\documentclass[letterpaper]{article} 
\usepackage{aaai2027}
\usepackage[hyphens]{url}  
\usepackage{graphicx} 
\usepackage{natbib}  
\usepackage{caption} 
\usepackage{algorithm}
\usepackage{algorithmic}

\usepackage{newfloat}
\usepackage{listings}
\DeclareCaptionStyle{ruled}{labelfont=normalfont,labelsep=colon,strut=off} 
\floatstyle{ruled}
\newfloat{listing}{tb}{lst}{}
\floatname{listing}{Listing}

\usepackage{booktabs}

\usepackage{amsmath,amsfonts}
\usepackage{array}
\usepackage{textcomp}
\usepackage{multirow} 
\usepackage{makecell}
\graphicspath{{Figures/}}

\title{NMKFR: A Robust Framework for Time-Aware Cold-Start Recommendation}

\author{
    Chengzhi Liu\equalcontrib\textsuperscript{\rm 1},
    Ning Zeng\equalcontrib\textsuperscript{\rm 1},
    Zehui Qu\textsuperscript{\rm 1}\corresponding,
}
\affiliations{
    \textsuperscript{\rm 1}School of Computer and Information Science, Southwest University, Chongqing, 400715, China\\

    \{liuchengzhi, zn18286222298\}@email.swu.edu.cn, quzehui@swu.edu.cn
}

\begin{document}
\maketitle

\begin{abstract}
Item cold-start recommendation is difficult when new items have sparse early interactions and appear in recommendation environments that keep changing over time. Static content, early feedback, and temporal-state evidence are all useful, but their reliability varies across the item lifecycle. This work proposes a framework—Neural Memory Kalman Fusion Recommender (NMKFR), which combines a Titans-based semantic encoder with time-aware Kalman state tracking. The semantic branch extracts memory-enhanced item observations from text, while the temporal branch estimates latent states under irregular interaction intervals. The NMKFR further uses posterior covariance as an uncertainty signal to calibrate semantic memory retrieval and adaptive static-temporal fusion. Experiments on Amazon Video Games and MovieLens-32M evaluate NMKFR under time-aware and item cold-start protocols using sampled candidate ranking. Across the reported comparisons, ablations, diagnostics, and robustness analyses, NMKFR achieves the strongest retained results and exhibits bounded uncertainty-related internal behavior. These findings provide empirical evidence for posterior-covariance-guided semantic–temporal fusion under the evaluated offline settings.
\end{abstract}


\section{Introduction}
\label{Introduction}
Recommender systems (RS) help users navigate large item catalogs by matching items with individual preferences. In modern platforms, these catalogs are not static collections: new items are launched, exposed, clicked, reviewed, and sometimes removed as user demand changes. Item cold-start appears at the beginning of this process, when a newly introduced item has little or no behavioral history and cannot yet benefit from the collaborative signals used for warm-start recommendation \cite{11176434, 3645494}. At this stage, the first ranking decisions shape both immediate recommendation quality and the feedback observed later, while the new item is still supported mainly by item-side content and sparse early behavior.

The evidence available for a new item changes over its early lifecycle. At launch, the ranking model mainly observes item-side fields such as text, metadata, categories, and tags. After limited exposure, a small amount of behavioral feedback may appear, but these interactions can be noisy because they are affected by initial exposure, short-term popularity, and incomplete user coverage \cite{11271180}. As the catalog continues to evolve, item attributes, user interests, and popularity patterns may also shift over time \cite{28721,7cdbd53d}. Static content, sparse early behavior, and temporal context therefore enter the ranking process with different levels of reliability at different stages of the item lifecycle.

Existing cold-start recommendation studies have improved how sparse item evidence is represented and supplemented. Collaborative and contrastive learning, meta-learning, graph and knowledge enhancement, and generative or large-model-based supervision all enrich new-item evidence when direct feedback is limited \cite{3583286,3475665,11176434,3657801,3531897,3703546}. These lines of work reduce missing information. After additional signals are obtained, the ranking model still has to compare evidence whose reliability changes with exposure and time.

\begin{figure}[t]
\centering
\includegraphics[width=0.95\linewidth]{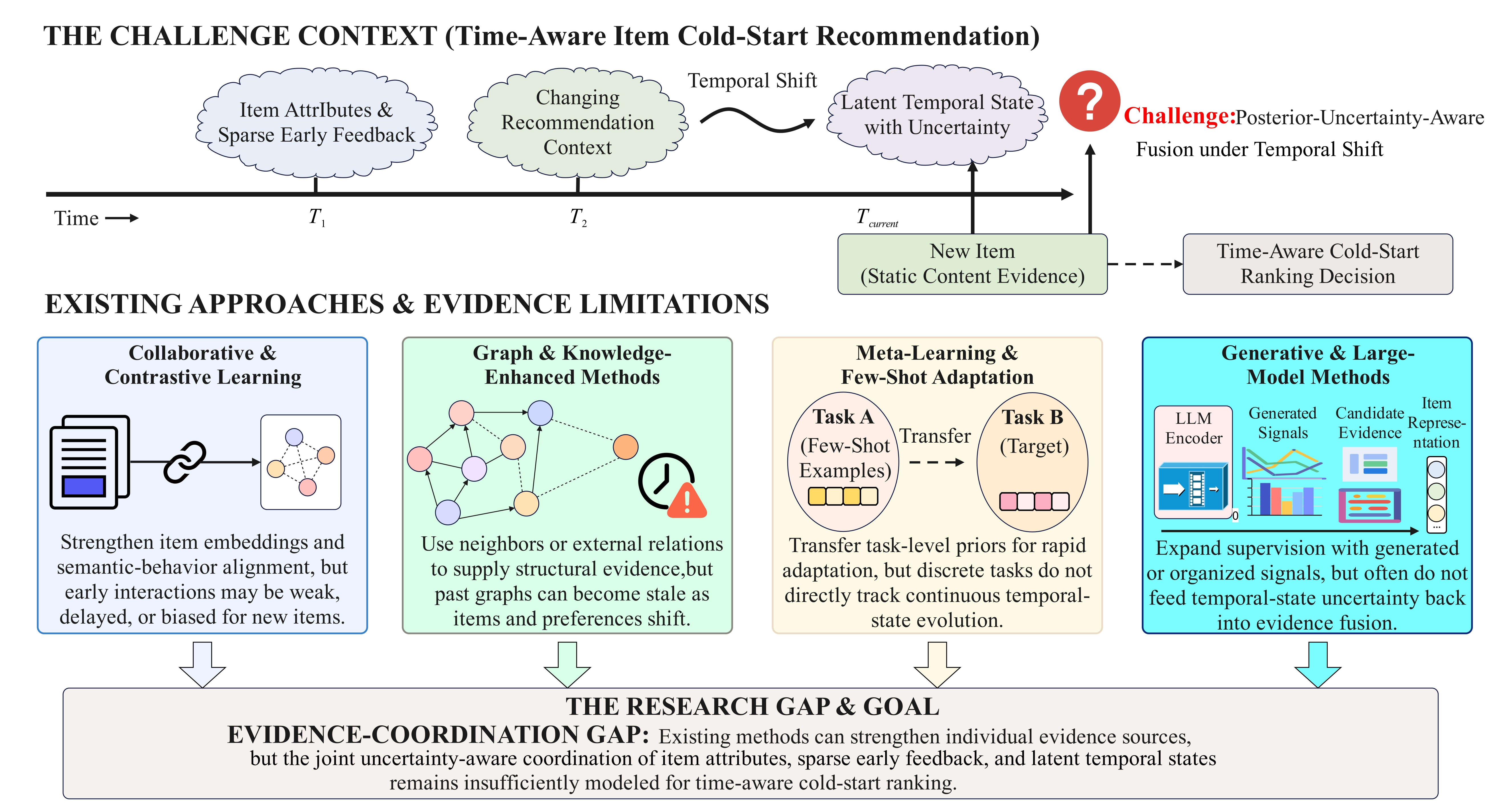}
\caption{Evidence-coordination gap in time-aware item cold-start recommendation.}
\label{fig:sceneGraph}
\end{figure}

When these strengthened signals are used for ranking, their reliability can change with exposure and time. Content features are observable early, but semantic similarity to warm items may fail to reflect current relevance after user interests or catalog composition shift. Early feedback can provide direct behavioral evidence, but it may also encode exposure bias, position effects, or short-lived interest before stable interaction patterns appear. Temporal signals are useful for tracking changes in popularity and user behavior, yet their estimates become less reliable when observations are sparse, irregular, or concentrated in a short period. Sequence and time-series models can encode temporal dependencies, but temporal encoding alone does not decide whether item-side evidence, early feedback, or temporal-state evidence should dominate a ranking decision under cold-start sparsity. Figure \ref{fig:sceneGraph} summarizes this evidence-coordination problem: existing methods can strengthen individual evidence sources, while the joint reliability of item attributes, early feedback, and latent temporal states remains insufficiently modeled.

A reliable ranking decision first needs a usable item-content observation. At the start of the item lifecycle, text is often the most stable item-side evidence because titles, descriptions, and tags expose attributes before reliable behavioral patterns have accumulated. The encoder should preserve local phrase-level evidence while retaining semantic cues that recur across token positions. Standard self-attention carries these signals through hidden states within a bounded context, whereas Titans separates bounded-context attention from neural long-term memory, giving cold-start text encoding local phrase matching and reusable semantic storage through associative reading and writing \cite{behrouz2025titans}.

Content observations still need to be interpreted at the timestamp where recommendation is made. The same textual evidence may be informative in one temporal context and less useful in another as popularity, exposure, and user interests change at different rates. A state-space formulation represents this context as a user-conditioned latent temporal state inferred from timestamped observations in the user's history. Kalman filtering supplies a prediction-update procedure, and a time-aware transition accounts for irregular intervals. The posterior mean gives a temporal-state representation, while the posterior covariance characterizes the uncertainty associated with the latent-state estimate.

Neural Memory Kalman Fusion Recommender (NMKFR) brings these operations into a single ranking pipeline. Feedback-Driven Titans Semantic Encoder (FD-TSE) transforms item text into a memory-enhanced semantic observation and, after the initial valid interaction, uses the preceding posterior covariance to modulate memory retrieval. Time-Aware Kalman Dynamics Tracker (TA-KDT) performs time-conditioned latent-state estimation from this observation. Uncertainty-Driven Feedback Mechanism (UDFM) converts the current posterior covariance into an uncertainty signal, and Adaptive Comparison Fusion Module (ACFM) uses this signal to adjust static-temporal fusion before ranking. The main contributions can be summarized as follows:

\begin{itemize}
\item A dual-branch recommendation framework is developed for cold-start recommendation under temporal shift. The semantic branch builds memory-enhanced item-content observations from text, and the temporal branch estimates time-conditioned latent states from item timestamps and intervals. The two branches retain separate static semantic and temporal-state representations before fusion.
\item A posterior-covariance-guided mechanism is introduced for static-temporal fusion. It converts the current posterior trace into an uncertainty signal that adjusts feature weighting in ACFM, while the preceding posterior covariance modulates FD-TSE memory retrieval after the initial valid interaction.
\item Experiments with 12 representative baselines separate ranking effectiveness from mechanism diagnostics. Overall, ablation and fixed-gain results evaluate NMKFR and its adaptive fusion, while Kalman diagnostics characterize the posterior signals used by UDFM.
\end{itemize}

\section{Related Work}
\textbf{Cold-Start Recommendation:} Cold-start recommendation has been studied through semantic enrichment, collaborative representation learning, limited-observation adaptation, relational propagation, generative supervision, and memory-based temporal modeling. These lines of work reduce missing evidence from different sides. Time-aware item cold-start further requires reliability coordination because item content, early feedback, and temporal context may not be equally trustworthy at the same ranking timestamp.

Cold-start recommendation connects static item attributes with likely behavioral responses under sparse feedback. Recent studies strengthen this connection through semantic-behavior alignment \cite{yao2025saviorrecsemant}, cross-modal sequential representation learning \cite{3657839}, temporal or distributionally robust optimization \cite{28721}, and models for user-interest drift, out-of-distribution cold-start, or stochastic temporal dynamics \cite{3592068,7cdbd53d}. Large language model simulators further provide auxiliary behavior signals when real interactions are limited \cite{3703546}. These methods enrich item evidence, while NMKFR focuses on how such evidence should be weighted when its reliability changes over the item lifecycle.

\textbf{Sparse-Evidence Representation and Adaptation:} Collaborative filtering and contrastive learning improve sparse item representations through auxiliary objectives, semantic-behavior alignment, co-occurrence calibration, preference aggregation, semantic null-space mapping, graph or diffusion augmentation, and alignment-uniformity regularization \cite{3583286,3475665,3657839,WangLiuShanSunChenGuan2024,3729894,11202202,LiWangXiaoYangZhouZhangJu2025}. These methods reduce representation degeneration, but their evidence can remain weak, delayed, or exposure-biased for newly introduced items \cite{11271180}.

Meta-learning methods reuse task-level priors for adaptation from limited observations \cite{11176434,3709336,WangPanZhaoLiangFengYao2025,3612446}. Cross-task collaboration, popularity-aware partitioning, enhanced user-preference estimation, augmented replay, hierarchical sparse-history structures, and uncertainty-aware stochastic meta-processes improve adaptation under sparse histories \cite{3709336,lietal2024shot,3612446}. Graph and knowledge methods aggregate high-order neighborhood information over user-item or knowledge graphs, simplify collaborative graph propagation, refine sparse graph structures, or incorporate continuous-time temporal graph signals \cite{3657801,10597916,3737026,3482242}. Generative and large-model-based methods produce cold-item representations, simulate behavior, augment training signals, or organize candidate evidence with language models \cite{3531897,3651532,3645494,3703546,zhang-etal-2025-llmtreerec,3717855}. These approaches supply additional evidence; NMKFR addresses how the ranking model coordinates static, behavioral, and temporal evidence after they are obtained.

\textbf{Neural Memory and State-Space Modeling:} Titans separates bounded contextual attention from neural long-term memory and improves long-range retention through memory update mechanisms \cite{behrouz2025titans}. This structure matches item descriptions, where local phrases, attribute lists, and cross-position semantic cues should be preserved before temporal-state estimation. Neural Kalman and probabilistic state-space hybrids show that learned representations can be combined with explicit state and uncertainty variables \cite{revach2021kalmannet,becker2024kalmamba}. NMKFR adapts this view to recommendation: item text supplies the semantic observation, and posterior covariance is fed back to semantic memory retrieval and static-temporal fusion.

\section{Method}

\begin{figure*}[t]
\centering
\includegraphics[width=0.95\linewidth]{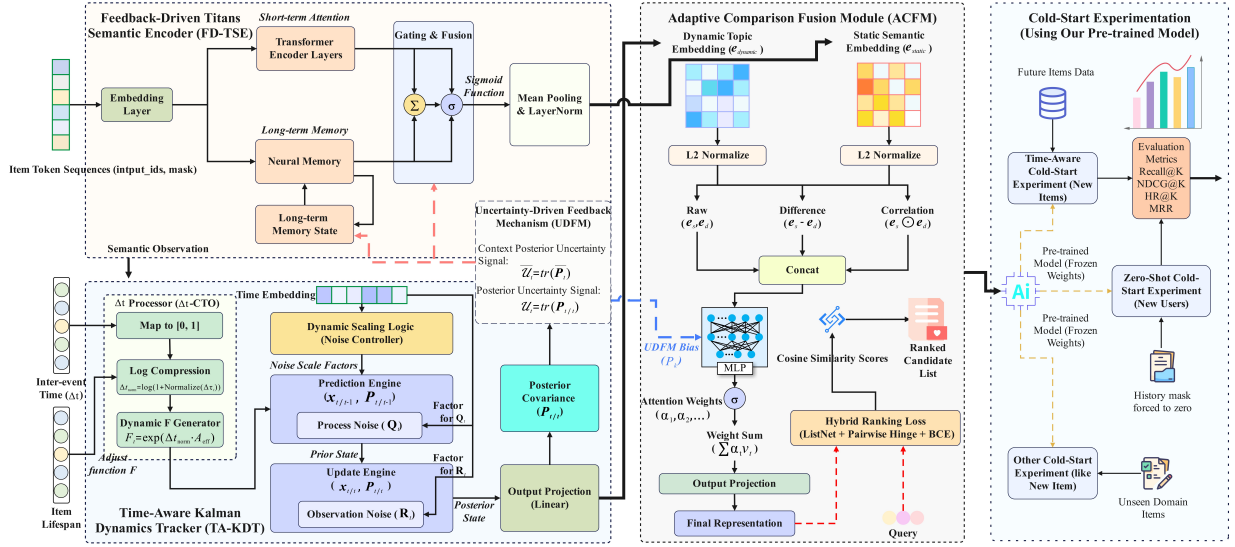}
\caption{Architecture of the proposed NMKFR framework. FD-TSE encodes item text, TA-KDT performs time-conditioned latent-state estimation, and UDFM supplies the current posterior uncertainty signal to ACFM for static-temporal fusion. After the initial valid interaction, the preceding posterior uncertainty signal modulates FD-TSE memory retrieval.}
\label{fig:model}
\end{figure*}

NMKFR ranks candidate items from item text and timestamped user histories. As shown in Figure \ref{fig:model}, FD-TSE encodes each history or candidate item as a semantic observation, TA-KDT applies a time-conditioned prediction--update operation to obtain its temporal-state representation and posterior covariance, and UDFM converts the current covariance into an ACFM uncertainty signal for ranking representation.

\subsection{Problem Formulation}
Let $\mathcal{I}$ be the item set. Item $v\in\mathcal{I}$ has text $T_v=\{w_1,\ldots,w_n\}$, and user $u$ has chronological history $S_u^{<t}=\{(v_1,\tau_1),\ldots,(v_{t-1},\tau_{t-1})\}$ before target time $\tau_t$. NMKFR learns a scoring function $f_\Theta(u,S_u^{<t},v,\tau_t)$ that ranks the observed next item against candidates available at $\tau_t$. Each item representation combines static semantic and time-conditioned temporal-state representations; the user representation pools valid history-item representations with a learned user embedding. The posterior covariance characterizes temporal-state uncertainty. Although the candidate may be cold, the user-conditioned latent state is propagated over $S_u^{<t}$ and summarizes timestamped history context rather than an item-specific interaction trajectory.

\subsection{Feedback-Driven Titans Semantic Encoder}
\label{FD-TSE}
FD-TSE maps item text to a semantic vector $h_t^{\mathrm{static}}\in\mathbb{R}^d$. It follows the MAG form of Titans, placing bounded-context attention and neural long-term memory in parallel so that local phrase evidence and reusable semantic patterns are modeled through separate paths \cite{behrouz2025titans}. Item descriptions are compact token sequences: product names, genres, attributes, and short descriptions can carry local cues, while repeated or long-span cues may describe more stable item properties. For token embeddings $X$, the short-term branch computes
\begin{equation}
        Y_{\mathrm{attn}}=\operatorname{TransformerEncoder}(X).
\end{equation}
This branch captures phrase-level contextual evidence within the current text window. In parallel, the memory branch keeps a differentiable read-write state across token positions. The key and value determine the association written into memory, while the query determines how stored information is retrieved for the current token:
\begin{equation}
\begin{aligned}
        k_j=W_K e_j,\quad v_j=W_V e_j,\quad q_j=W_Q e_j,\\
        M_j=M_{j-1}-\eta_t \nabla\ell_{\mathrm{mem}}(M_{j-1};k_j,v_j),\\
        y_{\mathrm{mem},j}=\mathcal{R}(M_j,q_j).
\end{aligned}
\end{equation}
The associative memory loss is written as
\begin{equation}
        \ell_{\mathrm{mem}}(M_j;k_j,v_j)
        = \frac{1}{2}\|\mathcal{R}(M_j;k_j)-v_j\|_2^2 .
        \label{eq:mem_loss}
\end{equation}
The attention path and memory path therefore provide complementary evidence before item-level pooling: one preserves local token interactions, and the other stores reusable semantic patterns across positions. The memory update is executed inside the FD-TSE forward pass, so ranking supervision can shape the Q/K/V projections, memory writing, retrieval, and final pooling jointly. For each noninitial valid interaction, FD-TSE uses the preceding posterior covariance, denoted by $\bar P_t$, and defines the corresponding uncertainty signal as $\bar{\mathcal{U}}_t=\operatorname{tr}(\bar P_t)$. The encoder uses this signal to modulate memory retrieval before the item-specific Kalman update:
\begin{equation}
        \hat y_{\mathrm{mem},j}=y_{\mathrm{mem},j}e^{-\bar{\mathcal{U}}_t}+e_j(1-e^{-\bar{\mathcal{U}}_t}),
        \label{eq:3}
\end{equation}
implemented with $\exp(-\operatorname{clamp}(\bar{\mathcal{U}}_t,10))$ for numerical stability. A larger context posterior uncertainty signal places more weight on the current token embedding. The attention and memory paths are then fused by a context-modulated gate,
\begin{equation}
        \alpha_t=\sigma(W_{gate}[Y_{\mathrm{attn}};\hat Y_{\mathrm{mem}}]-W_u\log(\bar{\mathcal{U}}_t+\epsilon)),
\end{equation}
\begin{equation}
\begin{aligned}
        \tilde H_t&=\alpha_tY_{\mathrm{attn}}+(1-\alpha_t)\hat Y_{\mathrm{mem}},\\
        h_t^{\mathrm{static}}&=\operatorname{MeanPool}(\tilde H_t,\mathrm{mask}).
\end{aligned}
\end{equation}
The context posterior uncertainty signal also biases the gate when it is available. For the first valid interaction, no preceding posterior covariance is available and FD-TSE uses the token-dependent attention-memory gate directly. At each subsequent valid interaction, the preceding posterior covariance provides the context signal for memory interpolation and gate biasing.

\subsection{Time-Aware Kalman Dynamics Tracker}
\label{TA-KDT}
TA-KDT treats the semantic vector as an observation of a latent temporal recommendation state. The static representation from FD-TSE is projected into the state observation space as $z_t=Mh_t^{\mathrm{static}}$, where $M\in\mathbb{R}^{k\times d}$ and $k=32$ in the experiments. Item text is thus a semantic measurement for time-conditioned latent-state estimation. To handle irregular intervals, the raw time gap is normalized and compressed,
\begin{equation}
        \Delta t_{\mathrm{norm}}=\operatorname{Log1p}(\operatorname{Normalize}(\tau_t-\tau_{t-1})),
\end{equation}
so intervals spanning different scales are mapped to a bounded temporal coordinate. The transition matrix is then obtained from a bounded matrix exponential:
\begin{equation}
        F_t=\exp(A_{\mathrm{eff}}\Delta t_{\mathrm{norm}}).
\end{equation}
This transition keeps the distinction between short and long gaps while avoiding an unbounded dependence on raw timestamps. The process and observation noise matrices use positive diagonal bases and bounded temporal scaling:
\begin{equation}
\begin{aligned}
        Q_t &= q_0s_t^Q I_k,\qquad R_t = r_0s_t^R I_k.
\end{aligned}
\label{eq:noise_param}
\end{equation}
Here, the bounded positive scales $s_t^Q$ and $s_t^R$ are derived from the temporal features at step $t$, with their parameterization given in the supplementary material. For each chronologically ordered user history, the latent state is initialized once with $x_{0|0}=\boldsymbol{0}$ and $P_{0|0}=I_k$. Each subsequent step propagates the preceding posterior state and covariance through the time-dependent transition:
\begin{equation}
        \hat x_t^-=F_tx_{t-1|t-1},\quad P_t^-=F_tP_{t-1|t-1}F_t^\top+Q_t.  
\end{equation}
Here, $F_t$ depends on the interval since the preceding valid interaction. Before the current semantic observation is incorporated, $\hat{x}_t^-$ and $P_t^-$ represent the propagated history state and its predicted uncertainty, respectively. The innovation $\nu_t$ measures the prediction--observation discrepancy, and the Kalman gain determines how strongly $z_t$ revises the predicted state:
\begin{equation}
\begin{aligned}
        \nu_t&=z_t-H\hat x_t^-,\\
        S_t&=HP_t^-H^\top+R_t+\epsilon I,\\
        K_t&=P_t^-H^\top S_t^{-1},\\
        x_{t|t}&=\hat x_t^-+K_t\nu_t,\\
        P_{t|t}&=(I-K_tH)P_t^-(I-K_tH)^\top+K_tR_tK_t^\top.
\end{aligned}
\end{equation}
The posterior mean is the time-conditioned temporal-state estimate, while the posterior covariance characterizes the remaining state uncertainty after observing $z_t$. The posterior mean is projected into the ranking space as
\begin{equation}
        h_t^{\mathrm{dynamic}}=W_{\mathrm{dyn}}x_{t|t}+b_{\mathrm{dyn}}.
\end{equation}

\subsection{Uncertainty-Driven Feedback Mechanism}
\label{UDFM}
UDFM uses posterior covariance to control the balance between retrieved semantic memory and temporal-state features. During the early cold-start phase and after larger temporal shifts, the two sources can carry different levels of state uncertainty. UDFM summarizes the current Kalman posterior covariance with an uncertainty signal:
\begin{equation}
        \mathcal{U}_t=\operatorname{tr}(P_{t|t}).
        \label{eq:13}
\end{equation}

The trace in Eq. \ref{eq:13} is the covariance summary used in the implementation. After TA-KDT processes the current semantic observation, UDFM applies $\log(\mathcal{U}_t+\epsilon)$ as a fusion bias in ACFM. FD-TSE uses a context posterior covariance $\bar P_t$ when a preceding valid interaction exists, and its trace $\bar{\mathcal{U}}_t=\operatorname{tr}(\bar P_t)$ controls memory interpolation through $\exp(-\operatorname{clamp}(\bar{\mathcal{U}}_t,10))$ before the item-specific state update. Thus, the current posterior uncertainty signal adjusts ACFM branch weights, while the preceding-posterior signal modulates FD-TSE memory access. Larger $\mathcal{U}_t$ increases the relative weight of static semantics, whereas larger $\bar{\mathcal{U}}_t$ favors the current token representation. During chronological propagation, the current posterior signal adjusts ACFM, while the preceding posterior covariance provides the context signal for FD-TSE at the next valid interaction.

\subsection{Adaptive Comparison Fusion Module \& Ranking}
\label{ACFM}
After UDFM produces the uncertainty signal, ACFM determines how static semantics $h^{\mathrm{static}}$ and temporal-state features $h^{\mathrm{dynamic}}$ should be combined for ranking. Simple concatenation or weighted summation does not explicitly compare agreement and disagreement between the two branches. ACFM therefore constructs three interaction features:
\begin{equation}
\begin{aligned}
        z_{\mathrm{cat}}&=[h^{\mathrm{static}}\|h^{\mathrm{dynamic}}],\\
        z_{\mathrm{diff}}&=|h^{\mathrm{static}}-h^{\mathrm{dynamic}}|,\\
        z_{\mathrm{corr}}&=h^{\mathrm{static}}\odot h^{\mathrm{dynamic}}.
\end{aligned}
\end{equation}
The concatenation term preserves the two branch representations, the absolute-difference term captures branch disagreement, and the element-wise product term captures agreement between static semantics and temporal-state features. These features are projected into two branch logits:
\begin{equation}
        \ell_t=W_1z_{\mathrm{cat}}+W_2z_{\mathrm{diff}}+W_3z_{\mathrm{corr}}+b.
\end{equation}
The posterior trace is then introduced as an uncertainty bias before softmax normalization:
\begin{equation}
        b_t^u=\lambda\log(\mathcal{U}_t+\epsilon),\quad
        \boldsymbol{\alpha}=\operatorname{Softmax}(\ell_t+[b_t^u,-b_t^u]).
\end{equation}
Where $\lambda>0$ is a learned scalar. The trace-dependent bias shifts the static and dynamic logits in opposite directions, with the learned magnitude of $\lambda$ determining the resulting branch preference.
\begin{equation}
        v_t=\alpha_{\mathrm{static}}h^{\mathrm{static}}+\alpha_{\mathrm{dynamic}}h^{\mathrm{dynamic}}.
\end{equation}
For user $u$, let $\mathcal{H}_u$ be the retained valid history items and let $v_i$ be the fused representation of history item $i$ produced by the preceding modules. The user representation $p_u$ and candidate score are
\begin{equation}
        p_u=e_u+\frac{1}{\max(1,|\mathcal{H}_u|)}\sum_{i\in\mathcal{H}_u}v_i,\qquad
        \hat y_{u,v}=\frac{p_u^\top v}{\|p_u\|_2\|v\|_2}.
\end{equation}
Here, $e_u$ is the learned user embedding and $v$ is the fused representation of a candidate item. For an empty retained history, the sum is zero and $p_u=e_u$. The cosine scores of all candidates form the ranking list used by the hybrid objective below.
To adapt to candidate ranking, NMKFR uses a hybrid objective that combines listwise, pairwise, and pointwise supervision:
\begin{equation}
        \mathcal{L}=\lambda_1\mathcal{L}_{\mathrm{ListNet}}+\lambda_2\mathcal{L}_{\mathrm{Pair}}+\lambda_3\mathcal{L}_{\mathrm{BCE}}.
\end{equation}
The combined objective trains NMKFR on both relative ranking order and point-wise relevance signals. In cold-start settings, early observations may not provide enough evidence for a single objective to learn stable ranking behavior.

\section{Experiment}
NMKFR is evaluated on Amazon Video Games and MovieLens-32M under the Time-aware Cold-Start and Item Cold-Start protocols. The experiments address four questions: \textbf{RQ1} compares NMKFR with representative baselines; \textbf{RQ2} studies component contributions; \textbf{RQ3} characterizes posterior dynamics across training and temporal gaps; and \textbf{RQ4} evaluates adaptive fusion under input noise and limited histories, together with its numerical response to synthetic temporal gaps.

\subsection{Experiment Setting}
\subsubsection{Datasets}
Two real-world datasets with different text length and catalog density are used: Amazon Video Games \cite{hou2024bridging} and MovieLens-32M \cite{2827872}. Amazon item text uses compact review-derived descriptions, while MovieLens item text combines titles and genres with user-generated tags. Detailed dataset statistics, precomputed item-text construction, preprocessing, and a separate timestamped semantic-evidence audit are provided in the supplementary material.

Both datasets use 70/10/20 chronological next-item splits with one positive and 100 timestamp-available negatives, excluding target and historical items. All models share splits, candidates, history restrictions, and metrics. \emph{Time-aware Cold-Start} evaluates complete validation and test streams, including training-seen and training-unseen targets, under timestamp-aware candidate availability. \emph{Item Cold-Start} retains only queries whose positive target has no training interaction. Histories retain up to 50 items; sequence lengths are 256 for Amazon Video Games and 96 for MovieLens-32M.

\subsubsection{Evaluation Metrics}
The evaluation reports Recall@K, NDCG@K, and MRR under the sampled-ranking protocol, where $K \in \{5,10,20\}$ for Recall and NDCG. Recall measures top-$K$ hit coverage, NDCG rewards higher-ranked positives, and MRR records the reciprocal rank of the first relevant item. Results are averaged over three runs; seed details appear in the supplementary material.

\subsubsection{Baselines}
NMKFR is compared with 12 representative baselines: CLCRec \cite{3475665}, CCFCRec \cite{3583286}, LightKG \cite{3737026}, M2GNN \cite{3591720}, PAML \cite{ijcai2021p0222}, SASRec \cite{8594844}, BERT4Rec \cite{3357895}, GLINT-RU \cite{3709304}, Mamba4Rec \cite{liu2024mamba4rec}, TDRO \cite{28721}, TPAB \cite{3709299}, and CDTR \cite{3645400}.

\begin{table*}[!t]
  \centering
  \small
  \setlength{\tabcolsep}{2pt}
    \begin{tabular}{c|l|ccccccc|ccccccc}
    \toprule
         &      & \multicolumn{7}{c}{Amazon Video Games }        & \multicolumn{7}{c}{MovieLens-32M} \\
    \midrule
    \textbf{Scenario} & \textbf{Model} & R@5 & R@10 & R@20 & N@5 & N@10 & N@20 & MRR  & R@5 & R@10 & R@20 & N@5 & N@10 & N@20 & MRR \\
    \midrule
    \multirow{13}{*}{\makecell{Time Aware\\Cold-Start}} & \textbf{BERT4Rec} & \underline{0.2883} & 0.3489 & 0.4134 & \underline{0.1956} & \underline{0.2352} & \underline{0.2615} & \underline{0.1962} & \underline{0.3127} & 0.4719 & 0.5348 & 0.2446 & 0.2789 & 0.3015 & 0.2041 \\
         & \textbf{CCFCRec} & 0.0904 & 0.1598 & 0.2831 & 0.0557 & 0.0779 & 0.1088 & 0.0769 & 0.0135 & 0.0246 & 0.0437 & 0.0081 & 0.0117 & 0.0165 & 0.0211 \\
         & \textbf{CDTR} & 0.0744 & 0.1391 & 0.2447 & 0.0445 & 0.0653 & 0.0917 & 0.0648 & 0.0651 & 0.1612 & 0.3451 & 0.0371 & 0.0676 & 0.1137 & 0.0677 \\
         & \textbf{CLCRec} & 0.1536 & 0.2548 & 0.4107 & 0.0984 & 0.1309 & 0.1702 & 0.1169 & 0.2857 & 0.4219 & 0.5123 & 0.1799 & 0.2242 & 0.2473 & 0.1773 \\
         & \textbf{GLINT-RU} & 0.2151 & 0.2946 & 0.3822 & 0.1515 & 0.1771 & 0.1992 & 0.1567 & 0.2119 & 0.2941 & 0.3533 & 0.1848 & 0.2115 & 0.2267 & 0.1474 \\
         & \textbf{LightKG} & 0.0958 & 0.1652 & 0.2837 & 0.0596 & 0.0818 & 0.1114 & 0.0798 & 0.2225 & 0.3117 & 0.4209 & 0.1524 & 0.1812 & 0.2087 & 0.1589 \\
         & \textbf{M2GNN} & 0.0436 & 0.0716 & 0.1135 & 0.0271 & 0.0364 & 0.0466 & 0.0395 & 0.2327 & 0.3145 & 0.4313 & 0.1495 & 0.1792 & 0.2086 & 0.1559 \\
         & \textbf{Mamba4Rec} & 0.1794 & 0.2585 & 0.3452 & 0.1188 & 0.1444 & 0.1662 & 0.1256 & 0.2779 & 0.3471 & 0.3927 & 0.1613 & 0.1831 & 0.1948 & 0.1371 \\
         & \textbf{PAML} & 0.1823 & 0.2675 & 0.3831 & 0.1241 & 0.1516 & 0.1807 & 0.1373 & 0.3036 & 0.4574 & 0.5022 & 0.2209 & 0.2562 & 0.2907 & 0.1952 \\
         & \textbf{TDRO} & 0.1412 & 0.2319 & 0.3612 & 0.0887 & 0.1172 & 0.1496 & 0.1044 & 0.2363 & 0.3324 & 0.4335 & 0.1587 & 0.1898 & 0.2153 & 0.1626 \\
         & \textbf{TPAB} & 0.0667 & 0.1179 & 0.2148 & 0.0412 & 0.0576 & 0.0818 & 0.0614 & 0.2102 & 0.3011 & 0.4113 & 0.1395 & 0.1687 & 0.1965 & 0.1469 \\
         & \textbf{SASRec} & 0.2249 & \underline{0.4166} & \underline{0.5522} & 0.1366 & 0.1981 & 0.2578 & 0.1583 & \underline{0.3127} & \underline{0.4759} & \underline{0.5438} & \underline{0.2462} & \underline{0.3278} & \underline{0.3425} & \underline{0.2681} \\
         & \textbf{NMKFR} & \textbf{0.2953} & \textbf{0.4256} & \textbf{0.5818} & \textbf{0.2008} & \textbf{0.2428} & \textbf{0.2822} & \textbf{0.2081} & \textbf{0.4142} & \textbf{0.4966} & \textbf{0.5685} & \textbf{0.3122} & \textbf{0.3389} & \textbf{0.3571} & \textbf{0.3014} \\
    \midrule
   \multirow{13}{*}{\makecell{Item\\Cold-Start}} & \textbf{BERT4Rec} & 0.1756 & 0.3093 & 0.4257 & 0.1238 & 0.1885 & 0.2319 & 0.1311 & 0.3355 & 0.4576 & 0.5724 & 0.1879 & 0.2583 & 0.2935 & 0.2098 \\
         & \textbf{CCFCRec} & 0.0572 & 0.1079 & 0.2074 & 0.0345 & 0.0507 & 0.0756 & 0.0561 & 0.0341 & 0.0621 & 0.1127 & 0.0207 & 0.0297 & 0.0423 & 0.0375 \\
         & \textbf{CDTR} & 0.1079 & 0.1809 & 0.2803 & 0.0657 & 0.0892 & 0.1141 & 0.0812 & 0.3428 & 0.4151 & 0.4697 & \underline{0.2508} & 0.2743 & 0.2882 & \underline{0.2409} \\
         & \textbf{CLCRec} & 0.1474 & 0.2726 & 0.4762 & 0.0899 & 0.1301 & 0.1812 & 0.1138 & \underline{0.3771} & \underline{0.5095} & \underline{0.6498} & 0.2228 & \underline{0.2936} & \underline{0.3258} & 0.2146 \\
         & \textbf{GLINT-RU} & 0.0145 & 0.0321 & 0.0533 & 0.0078 & 0.0134 & 0.0188 & 0.0205 & 0.0359 & 0.0404 & 0.0462 & 0.0252 & 0.0267 & 0.0282 & 0.0331 \\
         & \textbf{LightKG} & 0.0275 & 0.0675 & 0.1749 & 0.0152 & 0.0278 & 0.0545 & 0.0415 & 0.0312 & 0.0885 & 0.1371 & 0.0212 & 0.0389 & 0.1254 & 0.0606 \\
         & \textbf{M2GNN} & 0.0412 & 0.0683 & 0.1145 & 0.0253 & 0.0337 & 0.0391 & 0.0377 & 0.0091 & 0.0738 & 0.1736 & 0.0047 & 0.0247 & 0.1486 & 0.0563 \\
         & \textbf{Mamba4Rec} & 0.1678 & 0.2849 & 0.3965 & 0.1186 & 0.1747 & 0.2138 & 0.1293 & 0.2764 & 0.3627 & 0.4538 & 0.1903 & 0.2535 & 0.2947 & 0.1883 \\
         & \textbf{PAML} & 0.0549 & 0.1342 & 0.2554 & 0.0271 & 0.0525 & 0.0829 & 0.0522 & 0.2717 & 0.3543 & 0.4984 & 0.1865 & 0.2352 & 0.2789 & 0.1865 \\
         & \textbf{TDRO} & 0.0521 & 0.1114 & 0.2447 & 0.0314 & 0.0503 & 0.0833 & 0.0568 & 0.2875 & 0.4344 & 0.6053 & 0.1899 & 0.2371 & 0.2803 & 0.1981 \\
         & \textbf{TPAB} & 0.0467 & 0.0944 & 0.1929 & 0.0269 & 0.0421 & 0.0666 & 0.0491 & 0.0482 & 0.1278 & 0.2649 & 0.0023 & 0.0894 & 0.1672 & 0.0394 \\
         & \textbf{SASRec} & \underline{0.2017} & \underline{0.3538} & \underline{0.5271} & \underline{0.1316} & \underline{0.1967} & \underline{0.2368} & \underline{0.1582} & 0.3613 & 0.4728 & 0.6144 & 0.2137 & 0.2735 & 0.3017 & 0.2357 \\
         & \textbf{NMKFR} & \textbf{0.2456} & \textbf{0.3823} & \textbf{0.5546} & \textbf{0.1589} & \textbf{0.2028} & \textbf{0.2462} & \textbf{0.1712} & \textbf{0.3786} & \textbf{0.5393} & \textbf{0.6991} & \textbf{0.2592} & \textbf{0.3109} & \textbf{0.3514} & \textbf{0.2754} \\
    \bottomrule
    \end{tabular}%
  \caption{Overall comparison under the Time-aware Cold-Start and Item Cold-Start protocols on Amazon Video Games and MovieLens-32M. R@K denotes Recall@K and N@K denotes NDCG@K. The best results are in \textbf{bold}, and the second-best results are \underline{underlined}.}
  \label{tab:result1}%
\end{table*}

\begin{table}[t]
\centering
\scriptsize
\setlength{\tabcolsep}{3.4pt}
\renewcommand{\arraystretch}{0.80}
\begin{tabular}{llcccccc}
\toprule
\textbf{Variant} & \textbf{Name} & \multicolumn{3}{c}{\textbf{Amazon Video Games}} & \multicolumn{3}{c}{\textbf{MovieLens-32M}} \\
\cmidrule(lr){3-5}\cmidrule(lr){6-8}
 & & \textbf{R@10} & \textbf{N@10} & \textbf{MRR} & \textbf{R@10} & \textbf{N@10} & \textbf{MRR} \\
\midrule
w/o Kalman & NMKFR-S & 0.2263 & 0.1140 & 0.1047 & 0.4522 & 0.3013 & 0.2382 \\
w/o Titans & NMKFR-T & 0.2906 & 0.1465 & 0.1270 & 0.4538 & 0.2902 & 0.2242 \\
w/o UDFM & NMKFR-F & 0.2416 & 0.1239 & 0.1125 & 0.4378 & 0.2958 & 0.2357 \\
w/o ACFM & NMKFR-C & 0.1704 & 0.0819 & 0.0798 & 0.4418 & 0.2919 & 0.2298 \\
w/o $\Delta t$ & NMKFR-A & 0.2525 & 0.1302 & 0.1172 & 0.5560 & 0.3818 & 0.2504 \\
Full & NMKFR & \textbf{0.4410} & \textbf{0.2541} & \textbf{0.2173} & \textbf{0.5822} & \textbf{0.4066} & \textbf{0.3301} \\
\bottomrule
\end{tabular}
\caption{Component ablation results under a controlled ablation evaluation stream.}
\label{tab:ablation}
\end{table}%

\subsection{Overall Performance (RQ1)}
\label{result1}
Table \ref{tab:result1} reports the full per-baseline comparison under the Time-aware Cold-Start and Item Cold-Start protocols, using Recall@5/10/20, NDCG@5/10/20, and MRR. Across the two datasets and two protocols, NMKFR achieves the strongest overall performance among the reported results, covering contrastive, graph, meta-learning, sequential, state-space, and temporal-shift-aware baselines.

In time-aware cold-start, NMKFR improves over the strongest baseline by at least 2.16\% on Amazon Video Games and 3.39\% on MovieLens-32M. In item cold-start, the gains are at least 3.10\% and 0.40\%, with clearer MovieLens gains on ranking-sensitive metrics such as NDCG@10, NDCG@20, and MRR. The results support the use of memory-based item-text observations, time-conditioned state estimation, and posterior-covariance-guided fusion when interactions are sparse or temporally uneven.

\subsection{Ablation Study (RQ2)}
To assess component contributions, the full model is compared with five independently retrained variants: NMKFR-S removes Kalman tracking, NMKFR-T replaces Titans, NMKFR-F removes UDFM, NMKFR-C removes ACFM, and NMKFR-A removes explicit $\Delta t$ modeling. Within the controlled ablation stream, all variants use the same split files, 101-candidate construction, and evaluation metrics.

The full model obtains the best result across all ablation metrics. On Amazon Video Games, removing ACFM causes the largest NDCG@10 drop, which supports comparison-based static-temporal fusion. On MovieLens-32M, removing Titans most reduces NDCG@10 and MRR, while removing UDFM most affects Recall@10. The results support the complementary roles of time-conditioned state estimation, memory-enhanced text encoding, posterior-covariance-guided fusion, and irregular-interval modeling.

\subsection{Posterior Diagnostics and Temporal-Gap Analysis (RQ3)}
RQ3 characterizes the Kalman branch during optimization and across temporal-gap groups using the posterior covariance trace, an innovation-scale proxy, and bucket-level ranking quality. These diagnostics examine the numerical behavior of the signal supplied to UDFM, while its ranking contribution is evaluated by the ablation and fixed-gain comparisons.

\subsubsection{Optimization Dynamics and Posterior Stability}
To characterize TA-KDT during optimization, the diagnostic pass records the mean posterior covariance trace $\mathrm{tr}(P_{t|t})$ and the innovation-scale proxy $r_{\mathrm{global}}$ over trained checkpoints.

\begin{figure}[t]
\centering
\includegraphics[width=\linewidth]{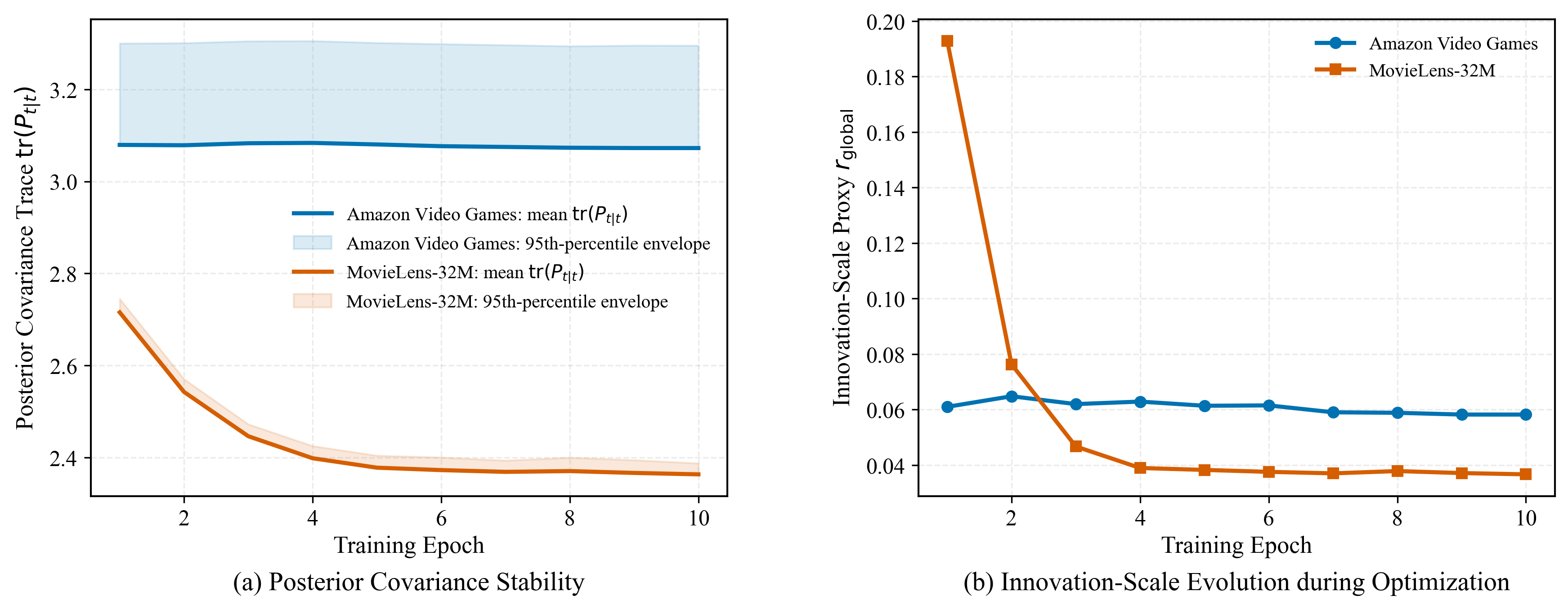}
\caption{Optimization trajectories of posterior covariance and the innovation-scale proxy.}
\label{fig:kalmant1t3}
\end{figure}

Figure \ref{fig:kalmant1t3} shows distinct but bounded optimization trajectories. On Amazon Video Games, the mean posterior trace and innovation-scale proxy remain nearly constant, while the non-widening 95th-percentile envelope indicates stable variation across observations. On MovieLens-32M, the innovation proxy decreases sharply before stabilizing, whereas the posterior trace declines more gradually. In both datasets, the trace remains finite without progressive dispersion or collapse, providing UDFM with a stable posterior-derived input.

\subsubsection{Temporal-Gap Stratification of Ranking and Posterior Trace}
\begin{figure}[t]
\centering
\includegraphics[width=\linewidth]{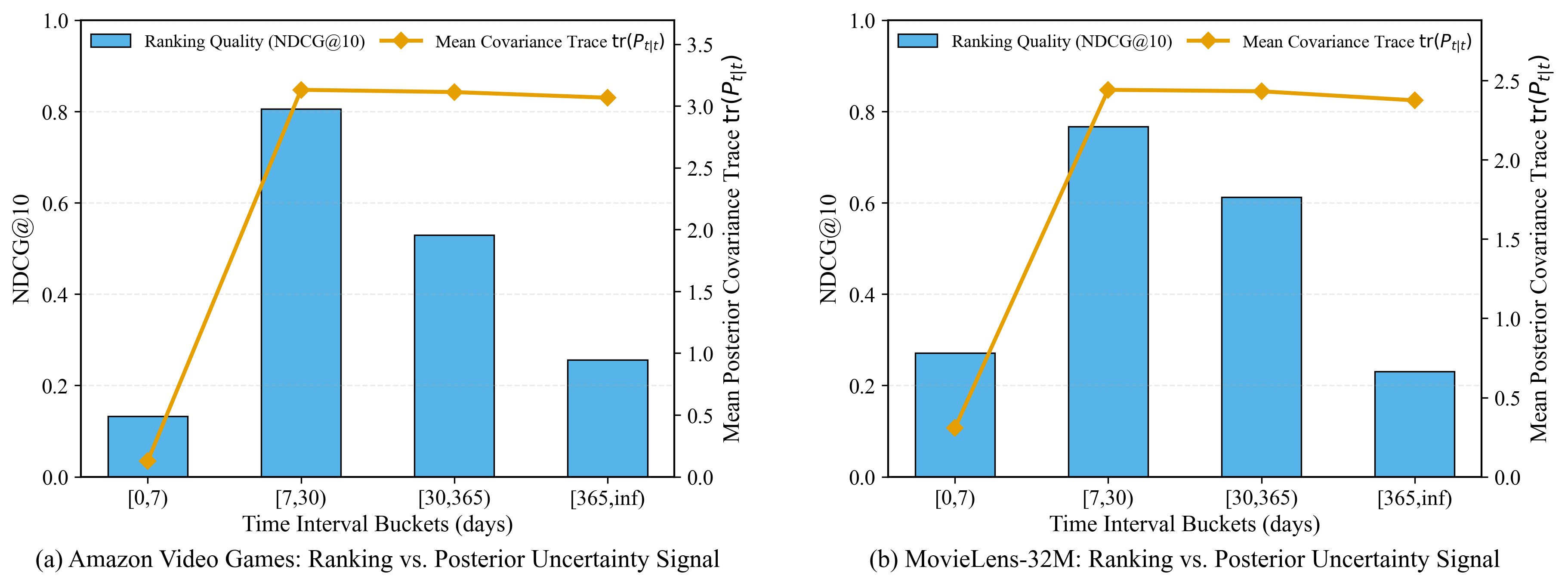}
\caption{Temporal-gap-stratified ranking quality and posterior covariance trace.}
\label{fig:kalmant2}
\end{figure}

Figure \ref{fig:kalmant2} groups candidates within cold-start ranking instances by their temporal gap and reports local NDCG@10 together with the average posterior trace. A query--bucket list contributes when it contains at least two candidates and one positive item.

Figure \ref{fig:kalmant2} shows that NDCG@10 peaks in the 7--30 day interval and decreases beyond 30 days, whereas the posterior trace remains bounded and varies more moderately. Thus, NDCG reflects bucket-level ranking difficulty, while the trace serves as a controlled input to adaptive static--temporal fusion. Its ranking utility is supported by the UDFM ablation and fixed-gain comparisons.

\subsection{Robustness Analysis (RQ4)}
\label{robustness}
RQ4 compares adaptive posterior-guided fusion with fixed-gain variants under token-level input noise and limited histories. A separate synthetic-gap experiment examines the numerical sensitivity of the Kalman quantities to controlled temporal inputs. The first two experiments report Amazon Video Games NDCG@10 and MovieLens-32M Recall@10.

\begin{figure}[t]
\centering
\includegraphics[width=\linewidth]{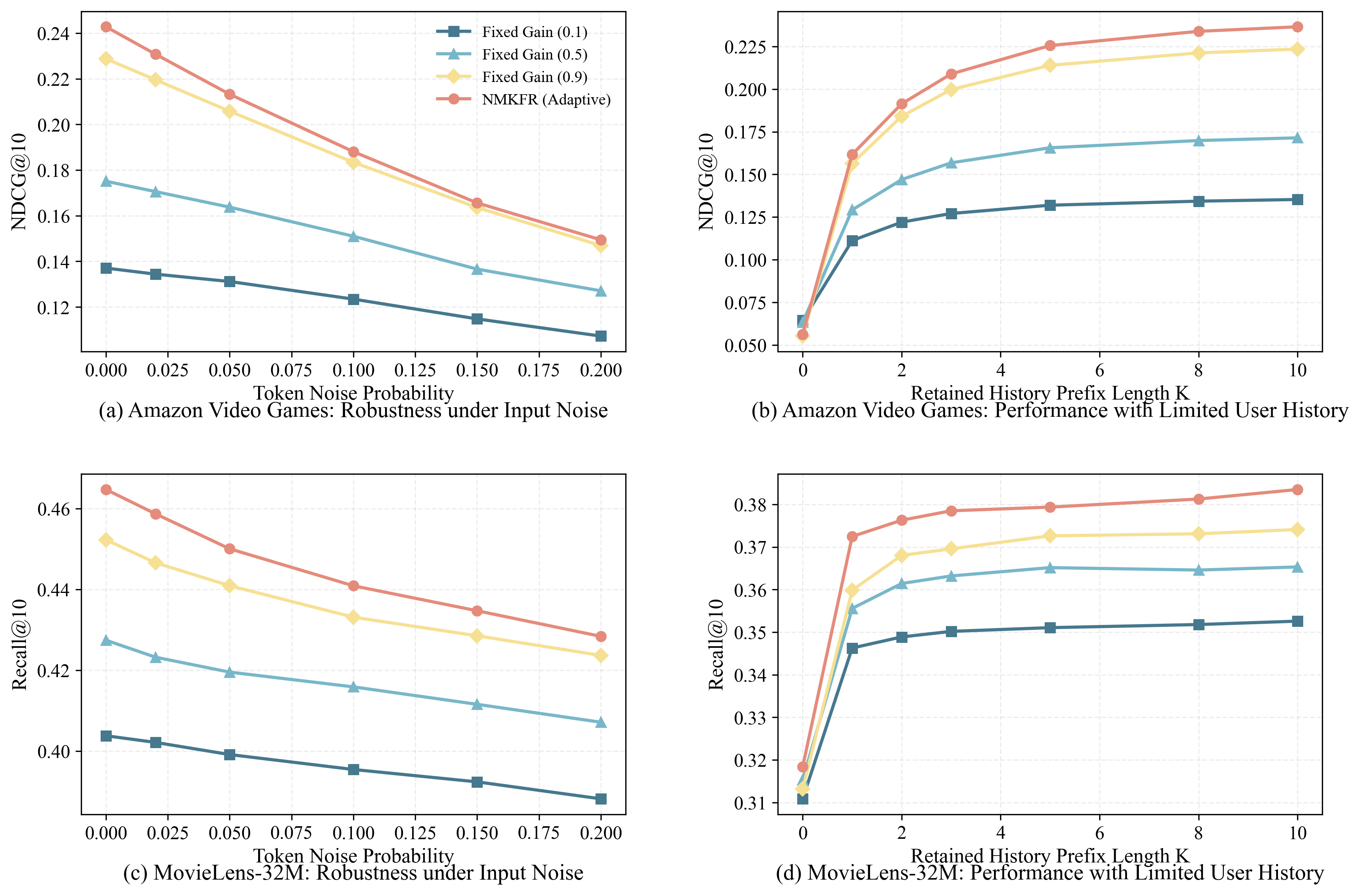}
\caption{Robustness analysis under token-level input noise and limited user history.}
\label{fig:analysis1}
\end{figure}

With token perturbation probability from 0.00 to 0.20, NMKFR remains strongest across the evaluated range in Figure \ref{fig:analysis1}. Amazon NDCG@10 decreases from 0.2428 to 0.1494, remaining above the strongest fixed high-gain result of 0.1469. MovieLens Recall@10 decreases from 0.4647 to 0.4284 while also remaining above the fixed-gain variants.

Across prefix lengths from 0 to 10, NMKFR leads at every tested point: Amazon NDCG@10 improves from 0.0562 to 0.2366, and MovieLens Recall@10 improves from 0.3184 to 0.3835. The fixed-gain variants remain lower or improve more slowly as additional interactions become available, supporting the use of adaptive fusion when the amount of historical evidence changes.

\begin{figure}[t]
\centering
\includegraphics[width=\linewidth]{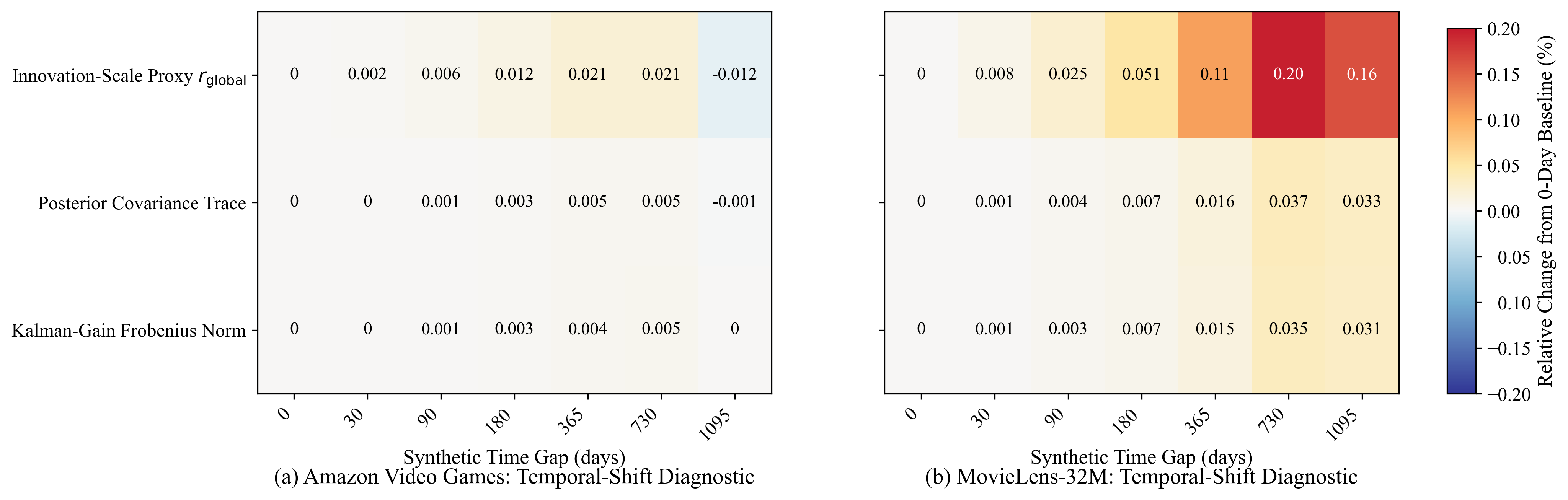}
\caption{Numerical sensitivity of Kalman diagnostics to synthetic temporal gaps.}
\label{fig:analysis2}
\end{figure}

In Figure \ref{fig:analysis2}, the synthetic-gap diagnostic varies the input interval from 0 to 1095 days. Relative to the zero-day reference, the maximum changes in the innovation proxy, posterior trace, and Kalman-gain norm are 0.021\%, 0.005\%, and 0.005\% on Amazon Video Games, and 0.20\%, 0.037\%, and 0.035\% on MovieLens-32M. These bounded changes show that the log-compressed interval representation and matrix-exponential transition remain numerically stable over the tested range. Ranking robustness is evaluated separately by the noise, limited-history, and ablation results.

\section{Conclusion and Future Work}
This paper presents NMKFR for item cold-start recommendation under temporal shift, integrating Titans-based semantic encoding, time-conditioned latent-state estimation, posterior-covariance-guided fusion, and ACFM. NMKFR obtains the best reported overall results across two datasets and the two evaluation protocols. The ablations support the contributions of its main components, while the fixed-gain comparisons support adaptive posterior-guided fusion under noisy inputs and limited histories. Posterior and synthetic-gap diagnostics further show bounded internal behavior over the evaluated range. These results support posterior-covariance-guided semantic--temporal fusion under the evaluated sampled-ranking protocols. Future work will study stricter feature-availability protocols, larger candidate evaluation settings, online feedback, and multi-modal item information.

\bibliography{aaai2027}

\end{document}